# Diameter-dependent thermopower of bismuth nanowires

Alla. Nikolaeva[1,2], Tito E. Huber[3], Dmitri Gitsu[1], and Leonid Konopko[1]

[1]*Institute of Applied Physics, Academy of Sciences of Moldova, Chisinau, MD-2028, Moldova.*

[2]*International Laboratory of High Magnetic Fields and Low Temperatures, Wroclav, Poland*[*]

[3]*Department of Chemistry, Howard University, 525 College St. NW, Washington, DC 20059.*

ABSTRACT

We present a study of electronic transport in individual Bi nanowires of large diameter relative to the Fermi wavelength. Measurements of the resistance and thermopower of intrinsic and Sn-doped Bi wires with various wire diameters, ranging from 150-480 nm, have been carried out over a wide range of temperatures (4-300 K) and magnetic fields (0-14 T). We find that the thermopower of intrinsic Bi wires in this diameter range is positive (type-p) below about 150 K, displaying a peak at around 40 K. In comparison, intrinsic bulk Bi is type-n. Magneto-thermopower effects due to the decrease of surface scattering when the cyclotron diameter is less than the wire diameter are demonstrated. The measurements are interpreted in terms of a model of diffusive thermopower, where the mobility limitations posed by hole-boundary scattering are much less severe than those due to electron-hole scattering.

# SECTION I. INTRODUCTION

One-dimensional materials, such as various kinds of nanowires (NWs) and nanotubes, have attracted considerable attention in theoretical and experimental studies.[1,2] Bismuth (Bi) is a semimetal with singular electronic transport properties that are related to its highly anisotropic Fermi surfaces, low carrier densities, small carrier effective masses, and large Fermi wavelength $\lambda_F$ (60 nm, as opposed to a few tenths of nm in most metals).[3] The small electron and hole Fermi energies of Bi result in a small carrier density ($3 \times 10^{17}/cm^3$ at low temperatures); nevertheless, Bi is a good conductor due to its high carrier mobility. Crystals of Bi are materials of high thermoelectric efficiency because of their small Fermi energies, which make the thermopower high, and also because of their intrinsic low phonon thermal conductivity.[4] The investigation of low dimensional thermoelectricity has recently become an active field, since electronic property modifications induced by confinement can be used to overcome the efficiency barriers imposed by the physics of bulk materials.[5,6] Therefore, the electronic transport properties and the thermopower of Bi NWs, specifically, are the subject of intense investigation.[7-16]

With the exception of temperatures below 10 K,[17-20] the thermopower of bulk Bi is due to the difference in broadening of the Fermi distribution between hot and cold regions of the sample or diffusion thermopower.[21-24] Because the mobility is larger for electrons than for holes, overall, in bulk samples, the thermopower of bulk Bi is negative. However, in low dimensional samples, the carrier's flight is restricted by boundary scattering, and the mobility is reduced accordingly. Therefore, the thermopower of Bi NWs can exhibit mobility size effects that have an origin in the unbalanced reduction of mobility of electrons and holes due to boundary scattering. Similar effects have been studied extensively in Bi thin films[25-31] and are known to be capable of changing the sign of the Hall effect. The effect of the reduction of mobility due to boundary scattering in the magnetoresistance of Bi NWs was studied.[10] However, systematic experimental studies of mobility size effects in the thermopower of NWs are scarce. Here, we focus on the study of the thermopower for NW diameters larger than 150 nm, for which quantum size effects can be disregarded and only mobility size effects are important. Theoretically, confinement causes the energies associated with transverse motion to be quantized, and the lowest energy level is increased, which results in an increase of the overlap energy between electron and hole bands $E_0$, 38 meV. For $d$ = 150 nm, one expects $\Delta E_o$ ~ 1-2 meV,[11] which can be safely neglected. For large diameter NWs, the expected effects should be even smaller.

Fabrication and electric measurements of bismuth NWs are challenging. The fabrication technique has to take into account the Bi oxidation instability, its low melting point, and its ease of alloying to materials that are typically used in making electrical contacts by soldering. Embedded NW arrays have been prepared in porous templates; however, electronic transport properties in such arrays are limited to 2-point resistance measurements with the number of wires measured being indeterminate. In addition, in thermopower measurements, the thermal conduction of the composite of alumina template and Bi NWs can cause experimental errors consisting of the overestimation of the temperature difference between the hot and cold side of the sample. We overcome these limitations by employing long Bi NWs prepared by the Ulitovsky method,[32] which employs glass-pulling techniques similar to those used in glass fiber-making to produce a single strand of Bi NW of controllable diameter in a glass fiber. Because the length of the NW employed in this study is a fraction of a millimeter, the resistance of the NWs is high, and contact resistance at the end of the NWs is within manageable limits. The strategy of using individual NWs for thermal and electrical transport measurements has been employed by other groups resolving high thermal conductivity issues with NW samples.[14,33-35]

An experimental study of the thermopower $\alpha$ of 240-620 nm-diameter individual Bi NWs has been presented.[36] It was found that $\alpha$ is positive at intermediate temperatures, below roughly 100 K, and negative at higher temperatures. Peaks of $\alpha$ of up to 80 µV/K observed at around 40 K are interpreted in terms of a model of diffusion thermopower under strong electron and hole-boundary scattering. The electron and hole density and the partial thermopowers were calculated using the Fermi energies obtained from the Shubnikov-de Haas (SdH) oscillations in the same samples. This study was performed for various elongations of the NW. The purpose of this report is to present an experimental study of resistance $R$ and $\alpha$ over a wider range of diameters and temperatures than in the study in Ref. 35, and to present a model of the electronic transport that takes these measurements into account. Another goal of this paper is to present measurements of $\alpha$ in the presence of a magnetic field. We observe that the thermopower becomes more negative for $B > B_C$, where $B_C$, Chamber's critical field for electron-boundary scattering avoidance, is $h\, k_F/\pi e d$. Here, $h$ is Planck's constant, $k_F$ is the carrier Fermi wavevector, and $e$ ( $=|e|$ ) is the electron charge.[37,38] In Bi NWs, Chamber's effects are signaled by an increase of electron mobility, and we observe a decrease of the positive thermopower, as expected from our model. We also present measurements of Sn-doped Bi NWs. Our model accounts for the experimental observations with doping, as well. Even though we restrict ourselves to the range of large diameter NWs, the phenomenon observed does not have to be restricted to large diameters; therefore, the insights that we gain in our study can be directly applicable to small diameter NWs of ($d \sim < \lambda_F$ ).

The paper is organized as follows: In Section II, we briefly describe the sample fabrication and characterization processes, as well as the experimental procedures for the transport measurements. In Section III, we present our measurements of $R$ and $\alpha$ at B=0, as well as a model of diffusion thermopower. In Section IV, we present the results of our measurements using the SdH method of partial thermopower of electrons and holes. In Section V, we present measurements of $\alpha$ of Sn-doped NWs and of NWs in the presence of an applied magnetic field parallel to the wirelength. Sections VI and VII conclude and summarize our paper.

**SECTION II. FABRICATION AND CHARACTERIZATION.**

The Ulitovsky method[32] for preparing individual Bi NWs is illustrated in the inset of Figure 1. This method, which involves casting from the liquid phase, consists of rapid drawing of a glass capillary from a glass tube containing the conductive melt. The purity of the Bi employed in this process is 99.999%. A prerequisite is that that the glass tube is maintained at high temperatures, sufficient to soften the glass. Under special conditions of temperature and for certain glass-metal combinations, the capillary contains a filament of the melt. The capillary is cooled with a water jet as it is drawn, resulting in cooling rates in excess of $10^6$ K/second. Segregation, which is problematic in the growth of Bi NW arrays of alloys such as Bi-Sb,[8] or Bi-Sn in the present study, is minimized due to the quick cooling. In NWs in the present study, the glass coating reliably protects the sample from mechanical damage while mounting and from oxidation. The technique employed yields extremely long (centimeter-length) monofilamentary samples, from which our millimeter-length samples are cut.

**Table. I.**  *Individual Bi NWs in a glass fiber: dimensions, compositions, and electrical resistance R at 300 K.*

| Sample | Electronic diameter $d$ (nm) | Capillary diameter $D$ (μm) | Sn concentration Atomic Percent | $R$(300 K) (KΩ/cm) |
|---|---|---|---|---|
| Bi-1 | 150 | 20 | 0 | 567 |
| Bi-2 | 240 | 24 | 0 | 252 |
| Bi-3 | 320 | 19 | 0 | 141 |
| Bi-4 | 480 | 22 | 0 | 63 |
| Bi-Sn | 200 | 24 | 0.02 | 428 |

A partial list of the samples employed in the work is presented in Table I. The samples are characterized by an electronic diameter $d = \sqrt{4L/\pi\sigma(300K)R(300K)}$, where $L$ is the wire length and $\sigma$(300 K) is the Bi room temperature conductivity of bulk Bi, which is[21] $8.7 \times 10^3 \Omega^{-1} cm^{-1}$. The actual diameter $d_{SEM}$ of representative samples was measured with a scanning electron microscope (SEM). A low-magnification SEM image of a 240-nm Bi NW is shown in Figure 1. High-magnification SEM images of the 240-nm and 480-nm NW cross-sections are presented in Figure 2.

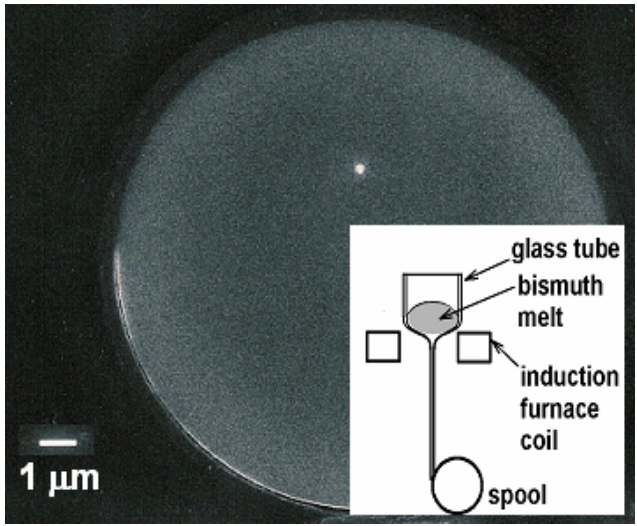

***Figure 1.*** *Scanning Electron Microscope cross-sections of the 240-nm Bi wire in its glass fiber envelope at the low magnification of 5000. The electron energy is 20 keV. The wire appears as light areas; the dark areas are the insulating envelope. The inset illustrates the Ulitovsky method for synthesizing long, small-diameter wires in a glass fiber.*

***Figure 2.*** *Scanning Electron Microscope cross-sections of 240-nm and 480-nm wires at high magnification. Dark backgrounds are insulating envelopes that extend out of the field of view. Bright areas are wires.*

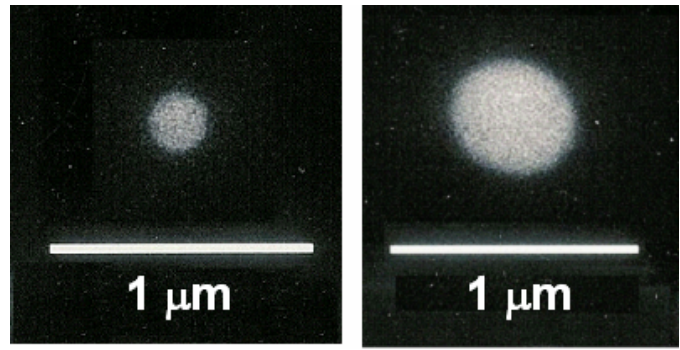

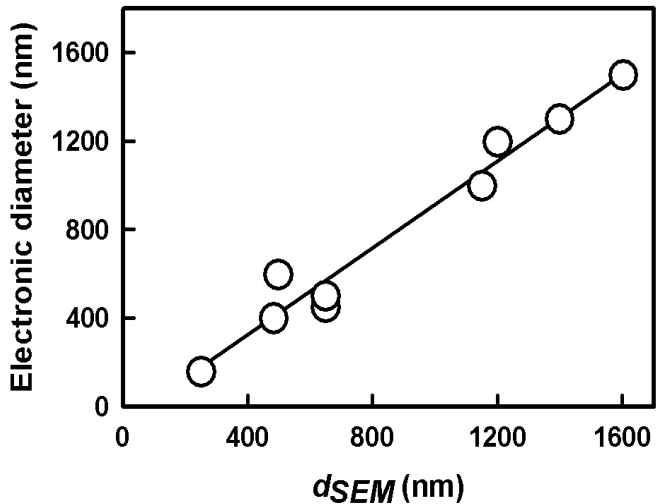

***Figure 3.*** *Electronic diameter d, determined by measuring the resistance of the wire and using Ohm's law, versus the diameter measured from SEM micrographs. The straight line is least-squares fit to the data.*

Figure 3 shows $d_{SEM}$ versus $d$; the error in $d_{SEM}$ is 5%. We find that $d_{SEM}= d$ within experimental error and that there is a trend toward $d_{SEM} > d$ for smaller diameters, consistent with resistance increases due to size effects at room temperature. Similar trends were observed by Toimil Molares and co-workers.[12] In this paper, we use $d$ to designate Bi NWs diameters rather than $d_{SEM}$.

Previous SdH studies have shown that the NWs grown by the Ulitovski method are single-crystal and that the orientation of the crystalline structure is such that the wirelength is in the bisectrix-trigonal plane making an angle of ≈70° with the trigonal axis $C_3$.[36] Figure 4 shows the orientation of our NWs with respect to the Fermi surface. The electron pockets $LB$ and $LC$ are located symmetrically with respect to the wire direction. To verify that the NW is a single crystal, we studied X-ray Laue diffraction patterns obtained using a cylindrical film camera (spindle was not rotating) and a monochromatic X-ray source. The exposure time was 90 hours. Sections of the same long NW had to be assembled in bundles in such a way that they were maintained parallel to each other. Clearly, it was important to keep the NWs from rotating around their axis during the assembly of the bundle. Preliminarily, the Laue patterns that we obtained were examined in comparison with well-known Laue diagrams for the As structure of Bi.[39] The examination shows a match with the diagram corresponding to As($10\bar{1}\bar{2}$), which is an orientation that is consistent with that derived from SdH studies.

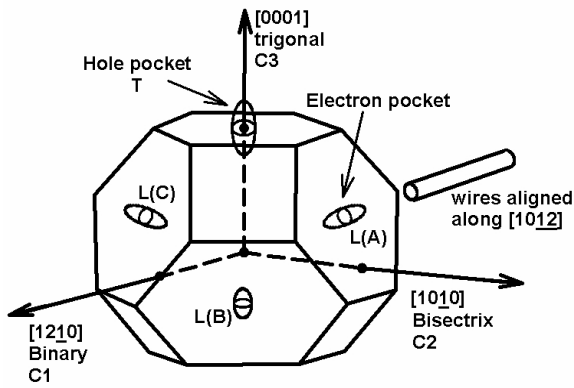

*Figure 4.* *The Brillouin zone for Bi, showing symmetry lines, which are indexed in the hexagonal system, and planes showing the orientation of the wires in the present study. The three Fermi surface electron pockets (LA, LB and LC) and the T hole pocket are also represented.*

The inset of Figure 5 shows the device employed for the resistance and thermopower measurements. This device was mounted in an appropriate cryostat. In the device, the NWs, which were 2-3 mm long, were mounted on a 5-mm-thick metal substrate. The wires were affixed to an epoxy substrate with electrodes. The heater allowed us to make a temperature difference between the 2 electrodes and therefore to the 2 ends of the wires. Electrical connections to the NWs were performed using a low-melting-point 58° C solder, In solder, or $In_{0.5}Ga_{0.5}$ eutectic. The latter type of solder makes very good contacts, but it has the disadvantage that it diffuses at room temperature into the Bi NW rather quickly. Consequently, the NW has to be used in the low temperature experiment immediately after the solder is applied. It is our experience that if the alloy is allowed to diffuse into the NW, the charge transport properties of the NW change; i.e. SdH oscillation is not observed. We measured the contact resistance between a 1-mm$^2$ contact of the solder and bulk Bi to be less than 10 μΩ at 4.2 K. Therefore, the contact resistance of a 160-nm Bi NW is estimated to be 390 Ω. However, we believe that the actual contact resistance is less than this estimate because: (a) The type of the solder employed did not affect the results of measurements, (b) there is no evidence of the superconducting transition of the solder at low temperatures (6 K for low melting point solder), and (c) the measurements of the resistance scales with the wire length.

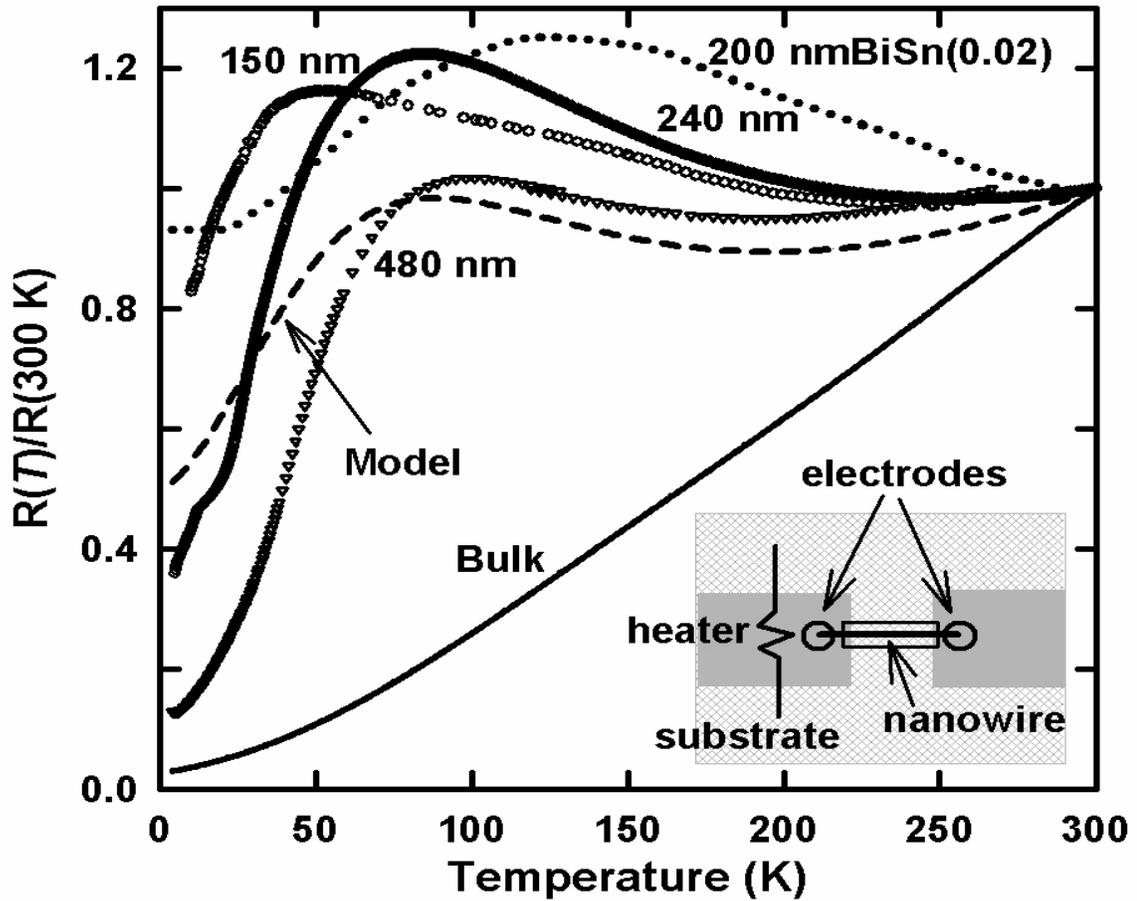

*Figure 5.*   *Temperature-dependent resistance for Bi and BiSn (0.02 at.%) nanowires of various diameters and of bulk polycrystalline samples, as indicated. The dashed line refers to the results of the model in Section IV with $\mu_{bound} = 10$ $m^2$ $V^{-1}$ $sec^{-1}$ and $v_b=50$ $m^2$ $V^{-1}$ $sec^{-1}$. Inset: Schematic of the arrangement used for the temperature- and magnetic field-dependent resistance and thermopower measurements. The device can be aligned with the magnetic field using a two-axis rotator that is available in the International High Magnetic Field and Low Temperatures Laboratory (IHMFLT), Wroclaw, Poland.*

## SECTION III. ELECTRONIC TRANSPORT (B=0)

Two types of cryostat with a base temperature of 4 K were employed. For the thermopower measurements, a Leybold 4.2 GM cryocooler was employed. The temperature difference between the cold end ($T_C$) and the hot end ($T_H$), $\Delta T_{HC}$, of the wire did not exceed 3°K. $\Delta T_{HC}$ was 1 K at low temperatures (i.e. 4 K). The magnetic field-dependent resistance $R(B)$, for magnetic fields $B$ of up to 14 Tesla (T) and for magnetic fields parallel to the wire length, was measured in a Bitter-type magnet and in a superconducting solenoid in the International High Magnetic Field and Low Temperatures Laboratory (IHMFLT, Wroclaw, Poland). The IHMFLT cryostat was equipped with a 2-axis rotator that facilitates the alignment of the NWs with respect to the magnetic field. Figure 5 shows the zero-field temperature-dependent resistance $R(T)$ of each sample, normalized to the resistance at 300 K. The figure includes results for diameters between 150 nm and 480 nm, and the corresponding values of the resistance at room temperature, $R(300\ K)$ are given in Table I. The resistance of a sample of bulk single crystal Bi from the literature[40] is also shown. The resistance ratio $RR = R(4.2\ K)/R(300\ K)$ of the NWs is a measure of carrier boundary scattering, and our samples had very small $RR$s compared to those reported for other samples of comparable diameter. The $RR$ the 240 nm sample in the present study is 0.35. In comparison, the $RR$ of the 200-nm wire array samples in Ref. 7 was 0.7.

The thermopower of Bi NWs of different diameters is shown in Figure 6. The thermopower of a single crystal of high purity Bi ($\Delta T // C_3$) was measured with the same method, and the results are also shown in the figure; our measurements agree with previous determinations of the thermopower of bulk Bi in this orientation.[21] The thermopower for $\Delta T \perp C_3$, from Ref. 21, is also shown in the figure. This is relevant here because, as shown in Figure 4, the orientation of our NWs is such that C3 is almost perpendicular to the wire length. There are two important features that are common to all the NWs that we have measured: $\alpha$ changes its sign from negative to positive between 100 K and 200 K, and it exhibits a peak at around 50 K. Also, we found that the value of the thermopower at room temperature was close to that of bulk Bi for the crystalline orientation of the NWs. Our measurements are not unlike the measurements of 200-nm Bi NW arrays by Heremans *et al,*[7] where the thermopower of 200-nm pure Bi NWs was found to be negative $T > 40$ K, and a small positive peak ($3 \pm 1$ μV/K) was observed at intermediate temperatures at around 20 K; however, in our measurements the maximum is much more pronounced.

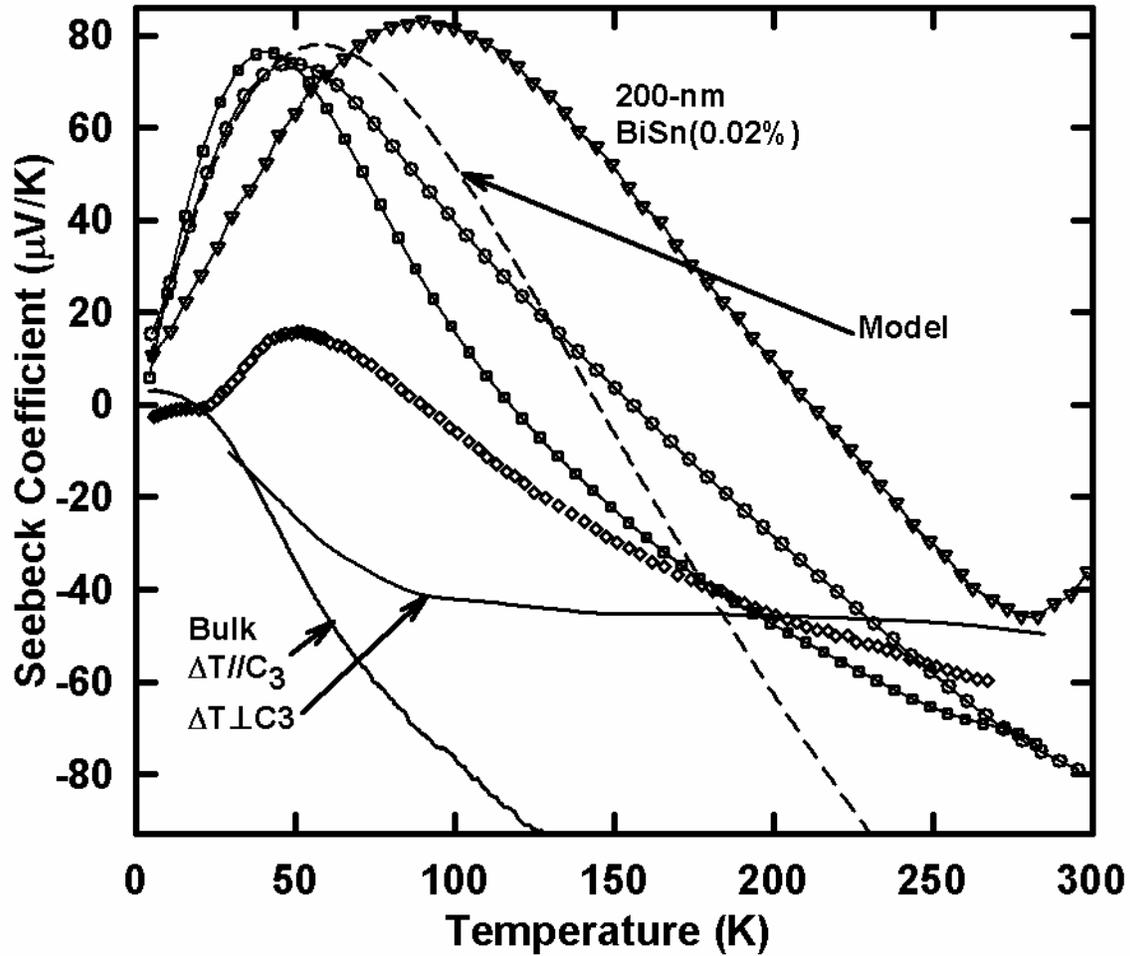

*Figure 6.* *Thermopower of individual Bi nanowires of various diameters. 150-nm, 240-nm, and 480 nm Bi NWs and 200-nm doped-Bi are indicated with open circles, squares, diamonds, and open triangles, respectively. The dashed line refers to the results of the model with $\mu_b = 10\ m^2\ V^{-1}\ sec^{-1}$ and $\nu_b = 50\ m^2\ V^{-1}\ sec^{-1}$. The thermopowers of bulk Bi crystals, orientation as indicated, are shown with solid lines.*

The phenomenological theory of low-field galvanometric effects in Bi in terms of carrier density and mobilities is derived from Boltzman equation[40] with Ohm's law holding for the resistivity tensor. This tensor must be invariant under the point symmetry operations of the crystal. This condition restricts the number of nonzero elements in the resistivity tensor to 6, and the number of independent and large components of the mobility tensor for electrons and holes to 3 and 2, respectively. $\mu_1$, $\mu_2$, and $\mu_3$ are electron mobilities, and $v_1$ and $v_2$ are the hole mobilities. The resistance, as well as the thermopower, of a macroscopic long cylinder of Bi obey the Thomson-Voigt relation.[21] For instance, $R = R_\perp \sin^2 \beta + R_{//} \cos^2 \beta$. Here $\beta$ is the angle between the cylinder axis and the trigonal direction; $\perp$ and // denote cylinder orientations perpendicular and parallel to the trigonal axis. It can therefore be expected that our case, since $\beta = 70°$ for Ulitovskii wires, corresponds approximately to the perpendicular case. From Ref. 40, we found that the partial conductivities of electrons $\sigma_e$ (=$1/\rho_e$) and holes $\sigma_h$ (=$1/\rho_h$), are:

$$\sigma_e = \frac{e}{2} n (\mu_1 + \mu_2) \qquad (1)$$

and

$$\sigma_h = e p v_1 \qquad (2).$$

where $n$ and $p$ are the electron and hole density, respectively, that we discussed in the previous section. Here $\rho = 1/(\sigma_e + \sigma_h)$.

In the diffusive case, we found that the thermopower is: an average of the partial thermopowers of electrons and holes, weighted according to the partial electric conductivities:[21]

$$\alpha = (\alpha_e \sigma_e + \alpha_h \sigma_h)/(\sigma_e + \sigma_h) \qquad (3)$$

where $\alpha_e$ and $\alpha_h$ are the partial thermopowers of the electrons and holes, and $\alpha_e < 0$ and $\alpha_h > 0$.

The electron and hole mobilities of bulk Bi have been determined.[40, 41] At 77 K, $\mu_1$=64 m$^2$V$^{-1}$ sec$^{-1}$ and $v_1$=10 m$^2$V$^{-1}$ sec$^{-1}$. $\mu_2$ is 2 orders of magnitude smaller than $\mu_1$. In bulk Bi, $\mu$ and $v$ increase for decreasing temperature as $T^{-1.5}$ achieving, at 4 K, values of $\mu_1=1.1\times10^4$ m²V⁻¹ sec⁻¹ and $\nu_1=2.2\times10^3$ m²V⁻¹ sec⁻¹. At that point, the mean free path (*mfp*) along certain directions was estimated to be 400 μm for electrons and 200 microns for holes. Since the diameter of the NWs studied in this work is much smaller than the *mfp* of the electrons and holes in Bi, especially at low temperatures, NW transport properties will experience size effects that originate in the boundary scattering of the carriers. The present section deals with the interpretation of resistance and thermopower size effects in terms of a simple model of mobilites that was found to be useful in the interpretation of Bi NWs magnetoresistance data.[6,10] Our interest in this model is that in it, surface effects can be discussed without specific reference to, but including, specular boundary scattering.[42] In this model, the carrier mobility has a mixed behavior; at high temperatures, the mobility increases for decreasing temperatures similarly to the bulk, and at intermediate temperatures the mobility saturates, becoming a constant at low temperatures. The reduction in the total carrier mobility by various scattering processes is as follows:

$$\mu_i^{-1} = \mu_i^{-1} + \mu_b^{-1} + \mu_{imp}^{-1} \tag{4}$$

$$\nu_1^{-1} = \nu_1^{-1} + \nu_b^{-1} + \nu_{imp}^{-1} \tag{5}$$

where $\mu_i$ and $\nu_i$ are the NW's electron and hole mobilities, respectively; $i=1$ and 2. $\mu_b$ and $\nu_b$ account for the boundary scattering of electrons and holes, respectively. $\mu_{imp}$ and $\nu_{imp}$ represent the effect of impurity scattering where $\mu_{imp} = \nu_{imp} = \infty$ for pure Bi NWs. In the following section, we use the model implicit in Eqs. 1-5 to fit the data.

### SECTION IV. ELECTRON AND HOLE PARTIAL THEMOPOWERS

In this section, we present the results of magnetotransport, including those using the SdH method, to determine the electron and hole Fermi energies. The Fermi energies are used to calculate the partial thermopowers. Figure 7 shows the low temperature longitudinal ($B // wirelength$) magnetoresistance $LMR = (R(B) - R(0))/R(B)$. A broad maximum, at $B_C$, is observed. These maxima can be interpreted in terms of the Chambers' effect.[37,38] According to Chambers, at low fields such that $B < B_C$, carrier scattering at the wire boundary dominates, and the resistance is high. As the field increases and the cyclotron diameter $d_c = h\, k_F/\pi eB$ is less than the wire diameter ($d_c < d$), boundary scattering becomes ineffective and the resistance decreases. The resistance of the NW decreases, approaching a limiting value of $R = L/(\sigma A)$, where $A = (\pi/4)d^2$ is the cross-sectional area of the NW and $\sigma$ is the conductivity of bulk Bi. The magnetoresistance is therefore negative, and the effect is more pronounced for high-mobility carriers. However, if the NW material has a small, but not zero, positive *LMR*, the resulting *LMR* shows a mixed behavior with a maximum occurring, very roughly, for a magnetic field $B_c$.

In practice, any misalignment results in a shift up of the *LMR* maximum, since the NWs have a very large transverse magnetoresistance. Therefore, to get the data for Figure 7, the NW was aligned with the rotator to minimize the value of $B_C$. The inset of Figure 7 shows $B_C$ as a function of diameter for the electron and hole pockets. A measurement of $B_C$ of 200-nm Bi wire arrays in Ref. 11 is shown for comparison. The lines in the inset represent $B_C = 2cp_F/ed$, where $p_F$ is $0.8, 1.1$ and $4.46 \times 10^{-21}$ g cm sec$^{-1}$ for electrons in the *LA*, *LB*, and *LC*-pockets and for holes, respectively.[3,43,44] (CGSE system). Therefore, we find good agreement between the maximum that is observed and $B_C$ for electrons. The decrease of the value of the resistance upon application of a large magnetic field is substantial. The 4 K resistance of 160-nm Bi NWs decreases by 80%, indicating a mobility increase of 5, from its zero-field value upon application of an 8 T field; in our case, $R$ has decreased to be only about twice the limiting value $R = L/(\sigma A)$. For holes, one expects a decrease of the LMR at $B_C$(holes)= 3.5 T in the 240-nm NW, but we found no evidence of a change of slope of the *LMR* for this particular magnetic field.

The NWs exhibit SdH oscillations that are due to the crossing of the Landau levels through the Fermi level.[45] These oscillations arise because, as the magnetic field is increased, the energy of the Landau level increases, and when its minimum value becomes equal to the Fermi energy, it is suddenly depopulated. The relaxation time for electron scattering is temporarily increased at this field value, giving rise to a dip in the magnetoresistance. The SdH oscillations are periodic in 1/B. The periods are equal to $(2\pi)^2 e / hcA_e$, where $A_e$ is the extremal cross-sectional area normal to the magnetic field of a pocket of the Fermi surface. Since the SdH periods measure the size of the pockets of the Fermi surface, they change dramatically if the Bi in the NW is doped with impurities. Since, in the process of making the sample molten, Bi comes into contact with glass that can potentially leech out impurities, it is possible that Bi is unintentionally doped with impurities. Doping with As, Te, and Sn has the effect of changing the Fermi level, upsetting the balance of concentrations of electrons and holes. We have therefore performed a careful study of the SdH periods in the NWs, for evidence of unintentional doping. As we show below, the periods observed correspond closely to those that are observed in pure crystalline Bi of the same orientation as our NWs, indicating that there is no unintentional doping.

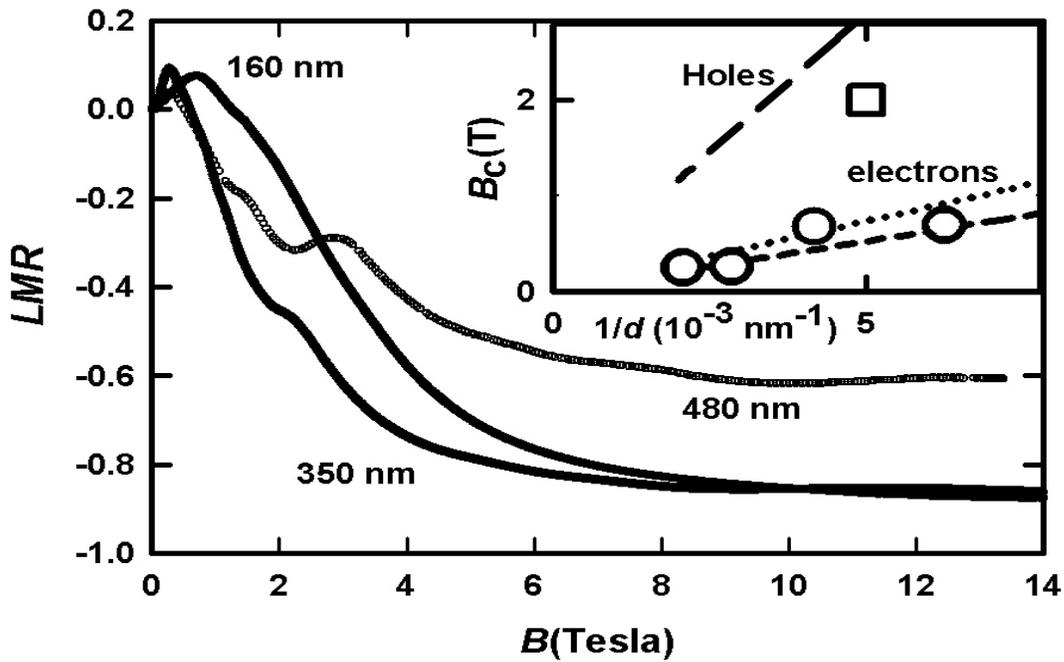

*Figure 7.* *Longitudinal magnetoresistance ($\vec{B} // \text{wirelength}$) as a function of the magnetic field for wires of various diameters, as indicated. The inset shows the experimental $B_c$ (empty circles) for the present work and an empty square for the data in Ref. 11. Calculated cut-off fields for holes (dashed line), for electrons in the LA-pocket (short-dashed line), and for electrons in LB- and LC-pockets (dotted line) are represented.*

Figure 7 shows that the *LMR* of the NWs in the present study is decorated with several peaks. These peaks are more clearly visible in the derivative of the *LMR*. Figure 8 shows the derivative of the *LMR*, as well as the maxima and minima positions, versus 1/*B* for our 480-nm Bi NWs. The maxima and minima are indicated with straight vertical lines. All the wires in the present study were investigated in this way, and we found that the value of the periods observed in pure Bi NWs is the same for all the NWs. These values are $P_T$ =(0.05±0.01) T$^{-1}$, $P_{LB,LC}$ =(0.34±0.05) T$^{-1}$ $P_{LA}$ =(0.85±0.05) T$^{-1}$. For the particular crystalline orientation of the NWs, and due to the relative magnitude of the periods, the pocket *LA* is denominated heavy electron and the pockets *LB* and *LC* are denominated light electrons.

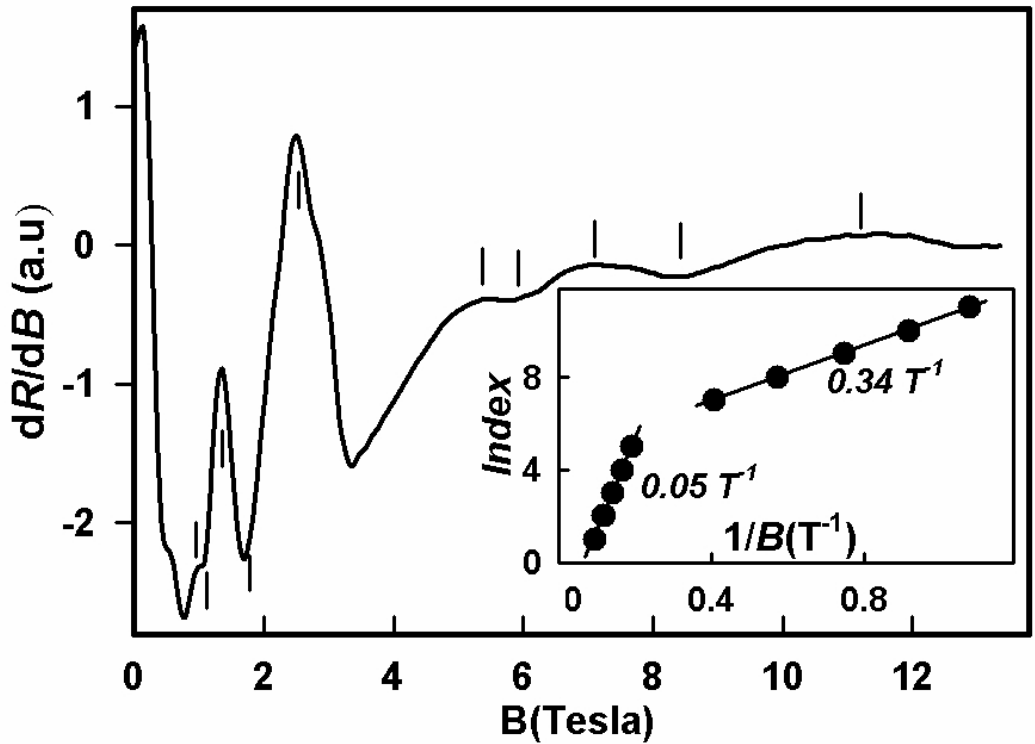

*Figure 8.*   *Derivative of the longitudinal magnetoresistance of the 480-nm individual Bi nanowire at T=4 K as a function of B. Vertical lines indicate the various maxima identified as arising from SdH effects. The inset shows positions of maxima and minima as a function of 1/B. SdH periods for holes (0.05 T$^{-1}$) and electrons (0.34 T$^{-1}$) are indicated.*

Within experimental error, the observed periods correspond closely to those found in bulk crystalline Bi for the orientation of the NWs.[46-48] The electron and hole periods depend sensitively upon the number of carriers, $n$ and $p$, in the Bi NWs, and therefore we find that, at low temperatures, $n = p = n_i$, where $n_i = 3 \times 10^{17}$ cm$^{-3}$ is the number of electrons or holes in intrinsic pure crystalline Bi. The hole number density for an ellipsoidal Fermi surface is given by the equation:[11]

$$p = (8\pi/3)\,(e/\pi hc)^{3/2}\,(P_{h,1}\,P_{h,2}\,P_{h,3})^{-1/2} \tag{6}$$

where $P_{h,1}$, $P_{h,2}$, and $P_{h,3}$ are the SdH periods with the magnetic field along the main directions of the ellipsoid. Considering that the ellipsoidal shape eccentricity does not change under doping, we find that the maximum level of doping in our NWs is $\Delta p/p_i = \pm 0.3$. This is relevant for our estimates of the thermopower in latter sections. SdH periods depend indirectly on the overlap and the Fermi energies, and therefore we find that $E_0 = 38$ meV and $E_F = 27$ meV in the NWs (same as in bulk Bi). The measurement precision is only fair, and since $P_h \sim (E_F^T)^{-1}$, where $E_F^T = E_0 - E_F$, and the error in the hole periods that are observed is found to be $\Delta P_h = 0.2\,P_h$, we find that the measurement error of the hole Fermi energy is $\Delta(E_0\text{-}E_F) = \pm 2.2$ meV. If this error is due to an impurity that comes inadvertently into Bi through processing, its effects would be felt more sensitively if the impurity is an acceptor or donor similar to that of Sn, Te, or Pb, and its effects would leave $E_o$, and ellipsoidal shape, invariant (this model is often called "rigid band" model in the literature). Therefore, we find that $\Delta(E_F) = \pm 2.2$ meV.

Askerov obtained the partial electron and hole thermopowers for a degenerate semiconductor in the Boltzman approximation for isotropic parabolic bands;[49] these are:

$$\alpha_e = -(k_B^2\pi^2 T/3e)\left[(r+1)(E_g^L + 2E_F)/E_F(E_g^L + E_F) - 4/(E_g^L + 2E_F)\right]$$

$$\alpha_h = (k_B^2\pi^2 T/3e)\left[(r+1)/E_F^T\right] \tag{7}$$

Here $r$ is the coefficient of the temperature-dependence of the phonon-part of the resistance. Although these expressions were derived for a semiconductor, they apply to a semimetal because the partial thermopower is a property of the bands, be it electron or hole (it does not depend upon $E_0$). The partial thermopower in a non-degenerate case, where $T$ is much less than the Fermi energy, is 0. The sign of the partial thermopower is related unambiguously to the sign of the charge of the carriers, and consequently the thermopower is instrumental in establishing the type of conductivity (type-*n* or type-*p*). Eq. (5) is derived in the isotropic approximation; however, Bi is anisotropic. We do not distinguish between the various pockets in Eq. (4). However, as a note, *LB* and *LC* have the highest mobilities and can be expected to contribute more than *LA* to the thermopower.

Heremans and Hansen[23] derived a different expression for $\alpha_L$ for the case where acoustic phonons dominate or, as in the case of the NWs, where scattering on the wall dominate.

$$\alpha_e = -(k_B^2 \pi^2 T / 3e) \frac{(E_g^L + 2E_F)}{E_F(E_g^L + E_F)} \qquad (8)$$

Taking the values appropriate for bulk Bi (see Table II), $r = \frac{1}{2}$, and using Eq. 7, we find $\alpha_e = (-0.81 \pm 0.1)\, T\, \mu V/K^2$ and $\alpha_h = (3.39 \pm 0.3)\, T\, \mu V/K^2$. According to Eq. (8), $\alpha_e = (-1.48 \pm 0.1)\, T\, \mu V/K^2$. The errors are related to the uncertainty regarding the Fermi energies. We have found the values in Eq. (7), rather than (8), to be appropriate for bulk Bi in our model.

*Table II.* Low temperature calculated carrier densities, Fermi energies, and partial and total thermopower for intrinsic and Sn-doped NWs at B=0.

| Sn at.% | 0 | 0.02 |
|---|---|---|
| $p$ ($10^{17}$ cm$^{-3}$) | 3 | 15±5 |
| $n$ ($10^{17}$ cm$^{-3}$) | 3 | not observed |
| $E_F$(meV) | 27 | not observed |
| $E_F^T$ (meV) | 11 | 47±5 |
| $E_0$(meV) | 38 | 38 |
| $\alpha_h$/T (Eq.1) ($\mu V/K^2$) | 3.4 | 1.1 |
| $\alpha_e$/T (Eq.1) ($\mu V/K^2$) | -0.8 | -5.2 |

We have observed that the transverse magnetoresistance *TMR =R(B)-R(0)* (*B* ⊥ *wirelength*) of the individual Bi NWs in the present study is a strong function of the orientation of the magnetic field. This dependence, denominated rotational diagram, is shown in Figure 9. We can explain the main features of our rotational diagram phenomenologically, as follows. As we indicated in Section III, the resistance and therefore the conductivity of crystalline Bi can be expressed as a function of the same quantities along and perpendicular to the trigonal axis. This function has the same angular symmetry as the experimental rotational diagram, and therefore the maximum and minimum of the experimental rotational diagram can be indexed with respect to C2 and C3, as indicated in the figure. The rotational diagram in Figure 9, in addition to the phenomenological angular dependence, has added structure that probably has a root in the details of the structure of the Fermi surface and would yield the SdH oscillations for each transverse magnetic field orientation, but this lengthy study is not presented here, since we focus on *LMR* oscillations.

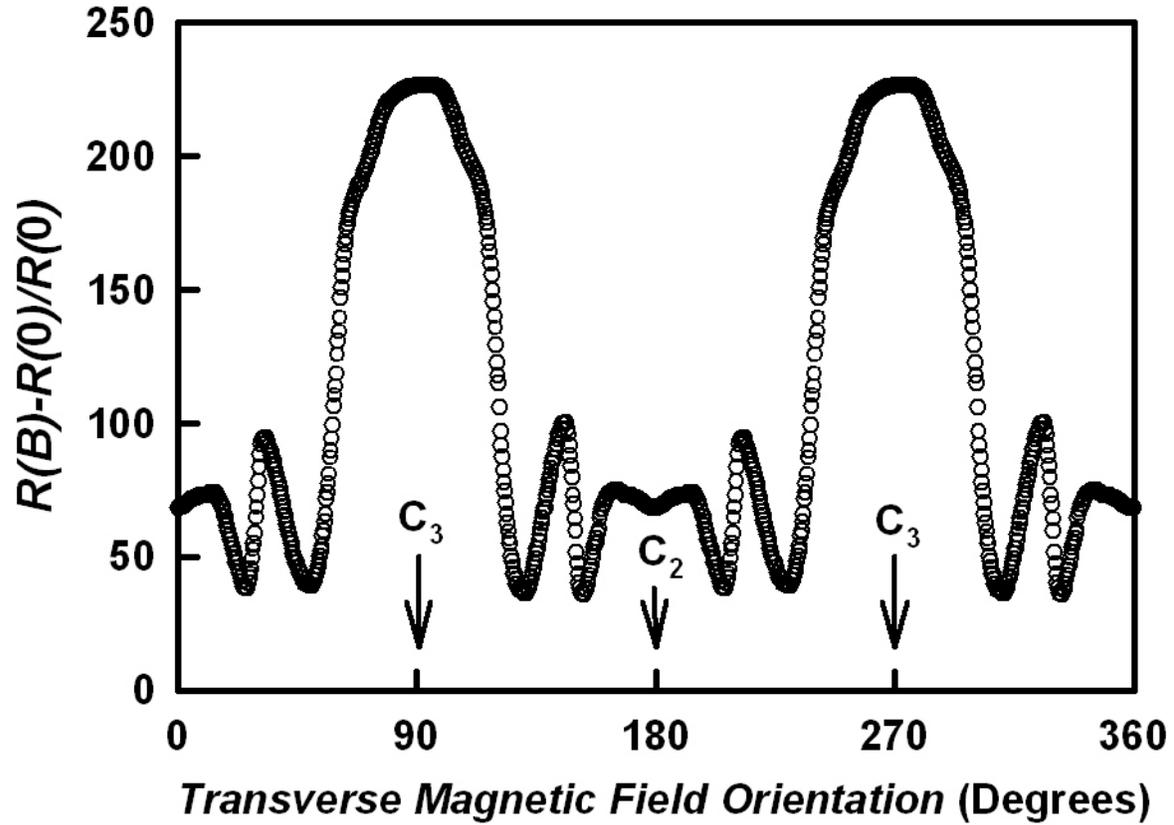

*Figure 9.* Rotational diagram of a 350-nm Bi NWs showing the dependence of the value of resistance at 4 K for fields of constant value and perpendicular to the wire length as a function of angle. $C_2$ and $C_3$ indicate angles for which $\vec{B}$ is aligned with the bisectrix and trigonal axis of the rhombohedral system.

The resistance and thermopower data for 240-nm wires in Figures 5 and 6 are fitted by $\mu_b$ = 10 $m^2 V^{-1} s^{-1}$ and $v_b$=50 $m^2 V^{-1} s^{-1}$. Figure 10 shows the temperature-dependent mobilities according to this model. At high temperatures, the bulk contribution to $\mu$ and $v$ becomes dominant, $\mu > v$, and the thermopower is negative and roughly equal to that of bulk Bi. Also, the resistance of the NWs is roughly equal to that calculated from the bulk Bi resistivity. This is observed in the experiments. While the model fit to the resistance of the 240-nm Bi nanowire is good, the model fits better the 150-nm data in the thermopower case; this anomaly can be corrected by decreasing the partial hole thermopower $\alpha_h$ by 7% which is within the margin of error of $\alpha_h$ that we are reporting. Since the results of our model (Eqs. 1-5) for Bi NWs indicate that the total mobility of holes is larger than that of electrons at low temperatures (T<80 K), the peak of the thermopower, at around 40 K, is associated with the interplay between bulk and surface contributions to the mobility. We note that the model reproduces the anomalous d$R$/d$T$ (d$R$/d$T$=0 at ~ 70 K for 240-nm NWs in Figure 5) of the temperature-dependent resistance.

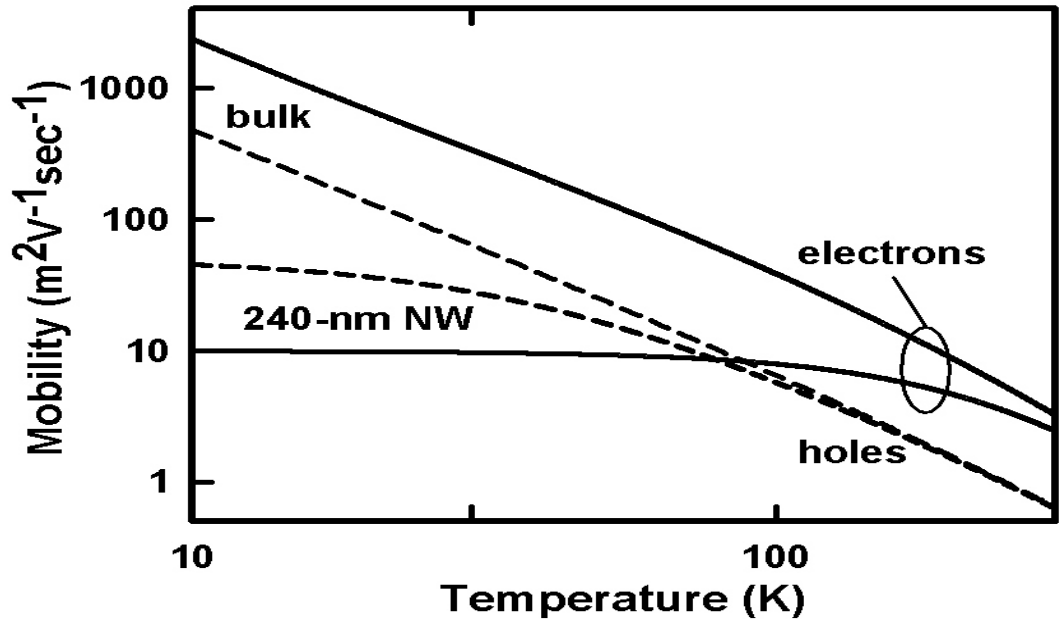

*Figure 10.*  *Temperature-dependent electron and hole mobilities in the model based on Eqs. 1-8. Electron $\mu_b$ = 10 $m^2 V^{-1} sec^{-1}$ and hole $v_b$=50 $m^2 V^{-1} sec^{-1}$. As indicated, solid lines are for the 240-nm NW, and dashed lines are for bulk Bi.*

## V. TIN DOPING AND B-DEPENDENT THERMOPOWER

We performed experiments involving doping of Bi NWs. The electron and hole concentrations $n_i$ and $p_i$ are small, and small additions of elements Sn, Te, and As from neighboring columns of the periodic table can produce large relative changes in carrier concentration,[50,51] thereby altering the transport properties. Preparing $Bi_{1-x}Sn_x$ solid solutions is not simple because Sn dissolves poorly in Bi, and it is known that this impurity is not distributed at random in Bi crystals; a superconducting transition, which occurs at very low temperatures, corresponding to Sn segregation is observed in Sn-doped Bi crystals samples pulled from the melt, as well as in zone refined samples.[52,53] However, the Ulitovsky method is uniquely suited for growing very homogeneous samples because the liquid-solid transition takes place at very high rates of cooling. Also, segregation involves only a small fraction of the Sn atoms in solution, and it is believed that segregations cause no adverse effects for T> 4 K. Our $Bi_{1-x}Sn_x$ NW main parameters are shown in Tables I and II. The resistance and thermopower of 200-nm with an Sn concentration of 0.02% are shown in Figure 5 and Figure 6, respectively. The Fermi energy of the holes was determined using the SdH method.[54] According to our SdH measurements, $E_F^T$ increased to about 47 meV in the NW samples (see Table II). Also, another effect of doping is the decrease of the mobility, due to the terms $\mu_{imp}$ and $\nu_{imp}$ in Eq. (4) and (5) that account for a broader $dR/dT=0$-anomaly in Figure 5 and broader thermopower maximum in Figure 6.

We also studied magnetic field effects on the thermopower. According to our model, the thermopower decreases with $B > B_C$ because this increases the electron's partial conductivity $\sigma_L$ and the partial thermopower of electrons is negative. The thermopower effect has been observed in a number of temperatures. Figure 11 shows the 55 K field dependence of the thermopower of the 240nm Bi wire. Hypothetically, the holes also have a Chambers' effect starting at $B_C$ (holes), at which point the trend toward negative values should reverse. However, there is no evidence of the holes' Chamber's effect in either the resistance, Figure 7, or the thermopower in Figure 11.

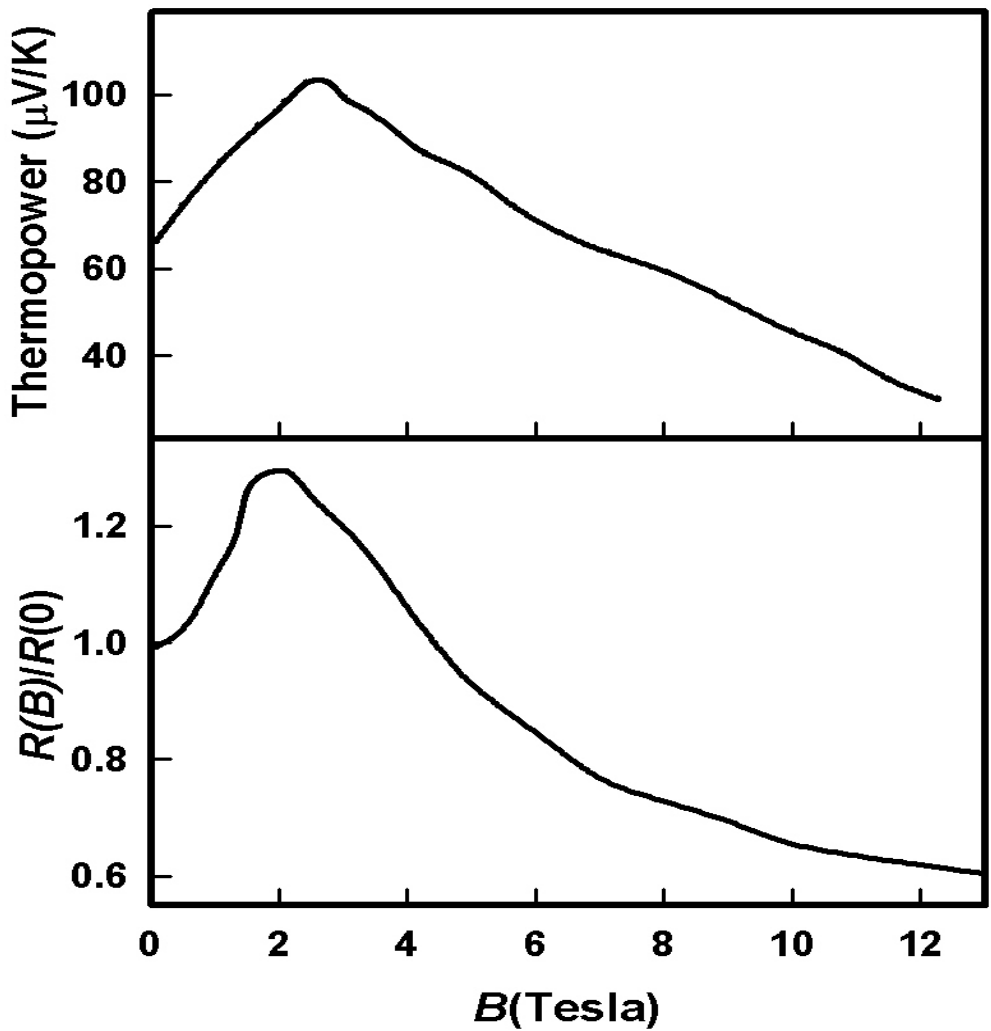

**Figure 11.** 55 K thermopower and normalized resistance of the 240-nm Bi NW versus longitudinal magnetic field.

We believe that the explanation for this negative result is that, for holes, this transition is spread out over a large range of magnetic fields and that the effect is not as noticeable for holes because they have high mobility. We have not performed a quantitative study of the experiments because the magnetic field, in addition to decreasing mobilities, changes $E_0$, and the analysis is outside of the scope of the present paper. These results will be presented elsewhere.

## SECTION VI. DISCUSSION

The finding of large mobilities in Bi nanowires is not surprising. In Ref. 10, for 70-nm wires, it was found that $\mu$=50 m$^2$V$^{-1}$ sec$^{-1}$ at low temperatures. The study of mobilities in terms of mean free paths reveals an interesting fact. Following Ref. 40, the electron relaxation time $\tau_e$ is $\mu_1 /[e\vec{m}^{-1}{}_{11}]$, where $\vec{m}^{-1}$ is the electron inverse effective mass tensor. Likewise, for holes, $\tau_h$ is $\nu_1 /[e\vec{M}^{-1}{}_{11}]$. Electron and hole mean free paths are calculated from relaxation times using $mfp_e = v_{F,e}\, \tau_e$ and $mfp_h = v_{F,h}\, \tau_h$, respectively. Here, $v_{F,e}$ and $v_{F,h}$ are the electron and hole Fermi velocities. Using these expressions, one finds that the low temperature mean free path of electrons and holes in bulk Bi is 400 μm and 200 μm, respectively. The mean free path of electrons and holes in the un-doped Bi nanowires in the present study can be estimated by considering that the Fermi surface parameters of such nanowires and that of bulk Bi are the same. Therefore we can use the values of inverse effective mass tensors and Fermi velocity given in Reference 40. For instance, for the 240-nm Bi nanowires, that has a hole mobility of 50 m$^2$V$^{-1}$ s$^{-1}$, we find $mfp_h$ =40 μm. What is surprising is that $mfp_h$ ~ 160 diameters. We believe that this effect is due to specular, or otherwise elastic, boundary scattering.

Although our thermopower measurements results are roughly consistent with those in Ref. 7, our interpretation with a diffusion thermopower model is a sharp departure, with an interpretation in terms of phonon drag offered in Ref. 7. Considering our margin of experimental errors and the simplicity of our model, our model fits simultaneously the thermopower data and the essential features of the resistance data, and indicates that mobility limitations posed by hole-boundary scattering are much less severe than those due to electron-hole. At low temperatures, where holes dominate the electronic transport, $\sigma(T) = v_b p$, where $p=n$ for pure Bi nanowires. From this equation, we can derive $v_b$ at 4 K, and the results are shown in Figure 12.

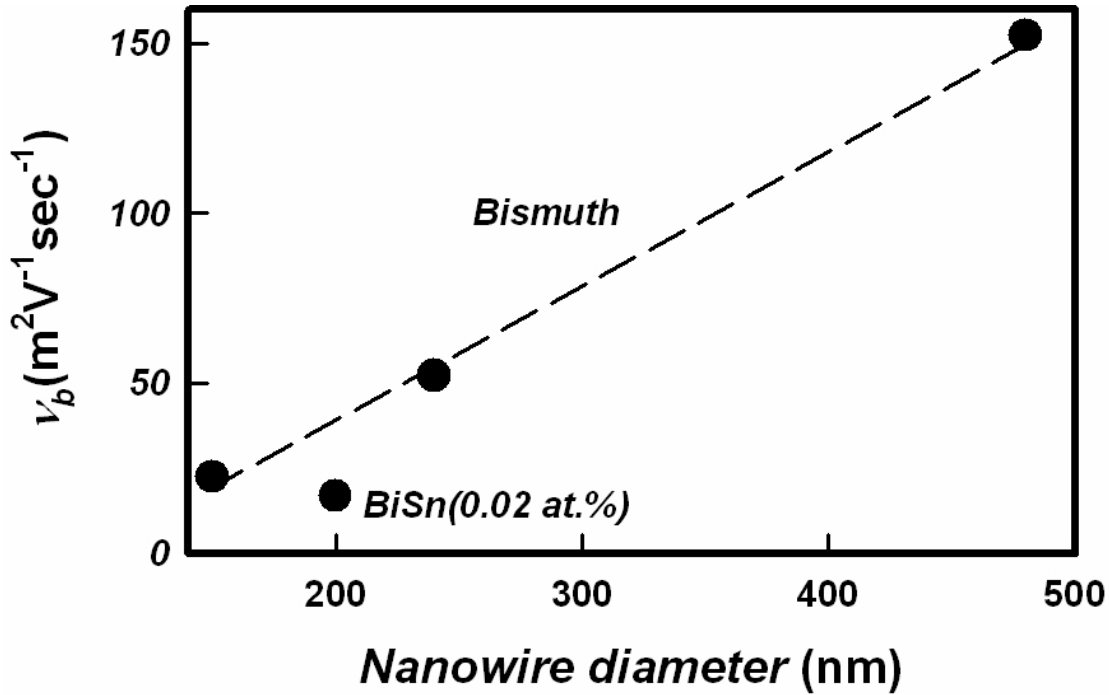

*Figure 12.* The low-temperature (4 K) mobility of holes in pure Bi and doped-Bi nanowires of various diameters. The dashed line is a guide to the eye.

Also, for temperatures below the thermopower maximum, since the carrier density is temperature independent, $\alpha = (\alpha_e \mu_b + \alpha_h v_b)/(\mu_b + v_b)$ and the thermopower is linear with temperature, as observed. Notice that we use the partial thermopower of electrons according to Eq. 7. Since the linear thermopower is approximately the same for all the NWs investigated, the conclusion can be reached that $\mu_b/\nu_b \ll 1$ for 150 nm $< d <$ 500 nm. The range of diameters, over which $\mu_b/\nu_b < 1$, may extend to 10-μm-diameter wires, as the results in Ref. 55 indicate.

The result that hole-boundary scattering is less than electron-boundary scattering ($\nu_b > \mu_b$) in confined Bi has been reported by other groups. One observation involves thin Bi films that were 20 nm to 500 nm thick and grown on CdTe insulator crystals by molecular-beam epitaxy.[29] Resistivity and Hall effect measurements were performed in a wide temperature range from 4 K to room temperature. The data were analyzed to extract electron and hole densities and mobilities from the field-dependent transport coefficients. At 100 K, it was found that $\nu/\mu \sim$ 1.3 and 1.5 for 500-nm thick films.[8]

In our model, the phenomenon of unbalanced reduction of mobility is a consequence of surface scattering and can be expected to become more relevant for NWs of smaller diameter ($d <$ 150 nm). Measurements of the thermopower of Bi NWs are scarce. Still, recently, a positive thermopower with a peak at 100 K has been observed in 55-nm individual polycrystalline Bi NWs that are reported to be free of impurities.[14] The observed effect may be associated with the phenomenon that is presented here, although the information available about the NWs is not complete enough.

The assessment of the significance of our results in the field of thermoelectrics, materials used for thermal to electric energy conversion, is as follows. Bi,[4] BiSb solid solutions,[56,57] and doped-$CsBi_2Te_3$ [58] are promising low-temperature thermoelectric materials at around 100 K. That these Bi-based materials have favorable combinations of high thermopower and electronic conductivity with low thermal resistivity is demonstrated by their high power factor $PF=\alpha^2\sigma$, and high thermoelectric efficiency $ZT=PF/\kappa$. Here $\kappa$ is the thermal conductivity. The results presented in previous sections can be used to evaluate the $PF$ and $ZT$ of Bi materials based on large-diameter (150 nm < $d$ < 500 nm) NWs, such as wire arrays. The power factor $PF$ can be calculated from our data, and we find that it peaks between 50 K and 100 K depending on the wire diameter and level of doping. The value of the $PF$ maximum can be very high; for example, for 240 nm NWs, $PF$ peaks at $6*10^{-5}$ W/cm*K$^2$, at 50 K, are about one order of magnitude higher than the $PF$ of bulk Bi at its peak at 100 K. The thermoelectric efficiency $ZT$ depends on $\kappa$, consisting of an electronic term $\kappa_{electr}$ and a phonon term $\kappa_{phonon}$, which has not been measured. We estimate $ZT$ as follows: Bi has a heavy ion that scatters phonons effectively, as shown in investigations of bulk Bi,[4] and theoretical and experimental studies indicating that $\kappa_{phonon}$ of Bi nanostructures is decreased from the bulk value.[2,59] If, indeed, $\kappa_{electr} > \kappa_{phonon}$, we can use the Wiedemann-Franz (WF) law: $\kappa_{electr} = L_0\sigma T$, where $L_0$ is the Lorenz number to estimate $ZT$. For example, for $\alpha = \sqrt{L_0} = 150\mu V/K$ one gets $ZT=1$. At 40 K and $B =2.3$ T, based on the value of $\alpha$ of 105 µV/K in Figure 11, one finds that $zT \sim 0.7$, more than twice the figure of merit of any other thermoelectric material at near 40 K, including doped-CsBi$_2$Te$_3$. There is another feature of Bi NWs that is of interest in the field of thermoelectrics. Most Bi-based thermoelectrics are type-n. The lack of suitable type-p materials is a major drawback for low-temperature thermoelectrics, which has turned the attention of device developers away from Bi thermocouples and into thermomagnetic devices that do not require materials of the 2 types. However, thermoelectric devices can be more efficient than thermomagnetic devices. In this context, the finding that surface scattering turns type-n materials to type-p NWs is very significant.

## VII. SUMMARY

This paper presents electrical transport and thermoelectric power measurements of long, individual single-crystal NWs of large diameter relative to the Fermi wavelength. Through measurements of the parameters of the electronic conduction bands, it was found that unintended doping of the NWs can be ruled out. A numerical simulation of the resistance and diffusive thermopower, considering bulk and surface boundary processes, was compared with the experimentally determined resistances and thermopowers, and good agreement was obtained. The similarities between experimental and simulated results strongly suggest that the observed effects are controlled by boundary scattering that limits the electron mobility more than the hole mobility. Furthermore, the magnetic field dependence of the thermopower and doping experiments are all consistent with the model being presented. We obtain the mobilities of electrons and holes. The mechanism for the scattering is not yet understood theoretically and needs further investigation.

The authors greatly acknowledge discussions with M.J. Graf. This work is supported by Civilian Research and Development Foundation for the Independent States of the Former Soviet Union (CRDF) # MP2 –3019. T.E.H. work was supported by the Division of Materials Research of the U.S. National Science Foundation under Grant No. NSF-0611595 and NSF-0506842.